\documentclass[journal,letterpaper]{IEEEtran}

\usepackage{graphicx,cite,epsfig,amssymb,amsmath,amsfonts,multicol,subfigure,mathtools,bm,mathrsfs,setspace}
\usepackage{multirow}

\newcommand{\tabincell}[2]{\begin{tabular}{@{}#1@{}}#2\end{tabular}}

\begin{document}
\title{\huge Cocktail BPSK: Achievable Data Rate beyond Channel Capacity}
\author{
	Bingli~Jiao,
	Yuli~Yang
	and~Mingxi~Yin
	\thanks{B. Jiao ({\em corresponding author}) and M. Yin are with the Department of Electronics and Peking University-Princeton University Joint Laboratory of Advanced Communications Research, Peking University, Beijing 100871, China (email: jiaobl@pku.edu.cn, yinmx@pku.edu.cn).}
	\thanks{Y. Yang is with the Department of Electronic and Electrical Engineering, University of Chester, Chester CH2 4NU, U.K. (e-mail: y.yang@chester.ac.uk).}
}

\maketitle

\begin{abstract}

In this paper, we propose a method layering two independent BPSK symbol streams in parallel transmission through a single additive white Gaussian noise (AWGN) channel and investigate its achievable data rate (ADR).  The receiver processes the two layered signals individually, where the critical point is to translate the interference between the two layers into the problem of one BPSK symbol having two equiprobable amplitudes.  After detecting one symbol stream, we subtract the detected result to demodulate the other symbol stream.   Based on the concept of mutual information, we formulate the ADRs and unearth an interesting phenomenon -- the sum ADR of the two BPSK symbol streams is higher than the channel capacity at low signal to noise ratio (SNR).  Our theoretical derivations together with illustrative numerical results substantiate this phenomenon.  We refer to this approach as \textit{cocktail BPSK} because the amplitudes of the two layered BPSK streams can be adjusted at the transmitter to achieve high spectral efficiency. 	
	
\end{abstract}


\IEEEpeerreviewmaketitle

\section{Introduction}

We consider the achievable data rate (ADR) of a parallel transmission composed of two independent BPSK symbol streams, where the amplitude of one stream is larger than that of the other.  This method is referred to as \textit{cocktail BPSK} since adjusting the energy intake proportion between the two streams plays a key role in achieving high spectral efficiency.  

The theoretical work starts with the formulation of input- and output signals over a memoryless additive white Gaussian noise (AWGN) channel described as 
\begin{eqnarray}
\begin{array}{l}\label{eq1}
y = x + n,
\end{array}
\end{eqnarray}
where $y$ is the output signal, $x$ is the input signal, and $n$ is the AWGN component from a normally distributed ensemble of power $\sigma_N^2$, denoted by $n \sim \mathcal{N}(0,\sigma_N^2)$.

To investigate the spectral efficiency of the finite-alphabet input, mutual information is adopted as a means of the ADR calculation, i.e., the maximum error-free transmission rate is calculated using~\cite{Shannon1948} 
\begin{equation}\label{eq2}
R = {\rm{I}}(X;Y) = {\rm{H}}(Y) - {\rm{H}}(N),
\end{equation}
where ${\rm{H}}(Y)$ is the entropy of the output signal and ${\rm{H}}(N) = {\log _2} (\sqrt{2 \pi e \sigma_N^2})$ is the entropy of the noise in (\ref{eq1}).

\begin{figure}[!t]
	\centering
	\includegraphics[width=0.45\textwidth]{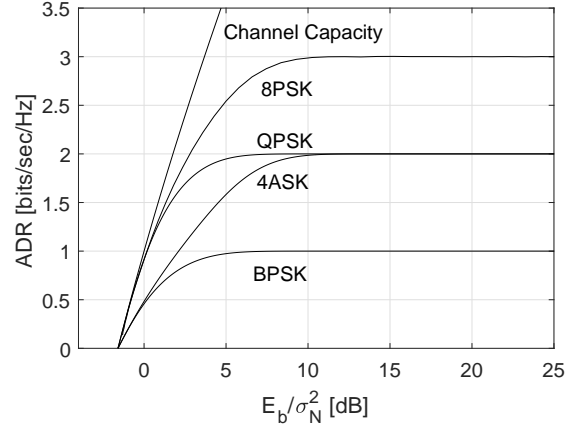}
	\caption{ADRs of BPSK, 4ASK, QPSK, 8PSK and the channel capacity.}
	\label{fig1}
\end{figure}

In the the previous literatures, the ADRs of popular modulation schemes, e.g., BPSK, QPSK, 8PSK and 4ASK have been calculated.  They are plotted in Fig. {\ref{fig1}} versus the logarithmic energy ratio of the bit to the noise in decibels, i.e., $E_b/{\sigma_N^2}$ in [dB].  To work mathematically on our derivations under study, we add an appendix to measure the ADR of BPSK versus the linear ratio of SNR. In addition, we find that the first derivative of the BPSK ADR equals to that of the channel capacity at the value of $\log_2 e$.  These results will be used in our theoretical derivations in the following.  

The mutual information of our proposed cocktail BPSK is calculated based on the following lemma: when the input $x$ in (1) consists of two independent signals, the ADR can be re-written by  
\begin{equation}\label{eq2}
R = \eta_1 {\rm{H}}(Y_1)+\eta_2 {\rm{H}}(Y_2) - {\rm{H}}(N),
\end{equation}
where $Y_i$ is the output sets pertaining to two independent input signals, and $\eta_i$  is the probability that $Y_i$ occurs, $i=1,2$ [1, page 53].  

As described in Shannon theory [1], maximizing the ADR with respect to the input distribution yields the capacity  
\begin{equation}\label{eqC}
C = \log_2 (1 + \rho) = \log_2 (1 + {\sigma_X^2}/{\sigma_N^2}),
\end{equation}
where the input signal $x$ is selected from a normally distributed assumable, i.e., $x \sim \mathcal{N}(0,\sigma_X^2)$ and, therefore, the SNR is ${\sigma_X^2}/{\sigma_N^2}$.

Inspired by the down-concavity feature of the channel capacity expression, i.e., $\log_2(1+\rho) \leq \log_2(1+\rho_1) + \log_2(1+\rho_2)$ when $\rho = \rho_1 + \rho_2$, we propose a parallel transmission concept, where two independent signal streams can be separated based on the use of both Euclidean and Hamming spaces to achieve higher data rates [6]. Since the two spaces are involved in the formulation, non explicit solution has been obtained due to extremely high complexity.

Therefore, this work pursues the thought of using parallel transmission to address the problem, again, by organizing the signals only with Euclidean geometry.  Consequently, we made the success for achieving high spectral efficiency beyond the channel capacity at low SNR.

The remainder of this paper is organized as follows.  Section II introduces the idea of cocktail BPSK and provides numerical results as the framework for analysis.  Section III provides theoretical analysis to further confirm the gain achieved by the proposed scheme. Finally, the paper is concluded in Section IV.

\section{Cocktail BPSK}
This section explains the communication mechanism of the cocktail BPSK, formulates ADRs based on the concept of mutual information and provides the illustrative numerical results.
\subsection{Communication Mechanism}
Consider the AWGN channel given in (1) with the input of two independent BPSK symbols in the parallel transmission.  In particular, the two BPSK symbols are denoted by $\alpha x_1$ and $\beta x_2$, where $\alpha$ and $\beta$ are the two amplitudes with the assumption of $\alpha>\beta>0$, and the symbols ${x_1, x_2}\in \{+1, -1\}$ with $+1$ and $-1$ occurring at the same probability.

The input signal, i.e., the signal of the cocktail BPSK, can be expressed by
\begin{eqnarray}
\begin{array}{l}\label{eqx}
x = \alpha x_1+\beta x_2.
\end{array}
\end{eqnarray}
Thus, the total input energy can be obtained statistically by
\begin{eqnarray}
\begin{array}{l}\label{eqR2}
E_{in}=\alpha^2+\beta^2,
\end{array}
\end{eqnarray}
which is deemed to be the symbol energy of the cocktail BPSK in the performance comparison with conventional modulation schemes.

The cocktail BPSK signal in (5) can be categorized into two cases: (I) when $x_1=x_2$ and (II) when $x_1=-x_2$.

Using (1) to detect $x_1$ yields 
\begin{eqnarray}
\begin{array}{l}\label{eqx}
y_1 = A_j x_1+n, \qquad j=1, 2,
\end{array}
\end{eqnarray}
where $y_1$ is the received signal pertain to the detection of $x_1$ and $j$ denotes the case index, i.e., $j = 1$ and $2$ pertain to Cases I and II, respectively.  Therefore, the amplitudes in these two cases, $A_1=\alpha+\beta$ and $A_2=\alpha-\beta$ occur at the same probability.  As it is very important to understand the two possible amplitudes of the symbol $x_1$, we present the details of Cases I and II in Table I.

\begin{table}[htb]
	\renewcommand{\arraystretch}{1.5}
	\centering
	%\tiny
	\small
	\caption{Parallel transmission of cocktail BPSK symbols.}
	\label{Table1}
	\begin{tabular}{c|c|c|c}
		\hline
		\tabincell{c}{Case} & \tabincell{c}{$x_1$} & \tabincell{c}{$x_2$}&\tabincell{c}{$y_1$} \\
		\hline		
		\multirow{2}*{I} & \tabincell{c}{$+1$} & \tabincell{c}{$+1$}& \tabincell{c}{$+(\alpha +\beta) +n$} \\
		\cline{2-3}
		~ & \tabincell{c}{$-1$} &\tabincell{c}{$-1$}  & \tabincell{c}{$-(\alpha+\beta)+n$} \\
		\hline		
		\multirow{2}*{II} & \tabincell{c}{$+1$} & \tabincell{c}{$-1$} &\tabincell{c}{$+(\alpha-\beta) +n$} \\
		\cline{2-3}
		~ & \tabincell{c}{$-1$}& \tabincell{c}{$+1$} &\tabincell{c}{$-(\alpha-\beta)+n$} \\
		\hline
		
	\end{tabular}
\end{table}

Upon the detection of $x_1$, we can recover $x_2$ by subtracting $x_1$ from $y_1$, i.e., 
\begin{eqnarray}
\begin{array}{l}\label{eqx}
y_2 = y_1-\alpha \hat x_1=\beta x_2 + n
\end{array}          
\end{eqnarray}  
where $y_2$ is used to detect the symbol $ x_2 $ and $\hat x_1$ is the recovered symbol based on the decision of detection of $x_1$. The last equality in (8) holds when $x_1$ is of error free transmission, which will be explained in next subsection.

\subsection{Achievable Data Rates}
In order to demonstrate the spectral efficiency of the cocktail BPSK, we calculate its ADRs in the following three steps.  

Firstly, taking the lemma of (3) into account, the ADR of $x_1$ can be calculated using   
\begin{eqnarray} \label{eqR1}
\begin{split} 
\mathbb{R}^{(1)} = \frac{1}{2}\mathbb{R}_\textrm{BPSK}(\gamma_1) + \frac{1}{2} \mathbb{R}_\textrm{BPSK}(\gamma_2),
\end{split}
\end{eqnarray} 
where $\gamma_1$ and $\gamma _2$ are the two SNRs pertaining to the amplitudes of $(\alpha+\beta)$ and $(\alpha-\beta)$, resectively, i.e.,$\gamma_1=(\alpha+\beta)^2/ \sigma_N^2$ and $\gamma_2=(\alpha-\beta)^2/ \sigma_N^2$.  Moreover, $\mathbb{R}_\textrm{BPSK}(.)$ is the ADR of conventional BPSK modulation, detailed in the appendix.

Secondly, since $x_1$ can be of error-free transmission theoretically with the up-bound of (9), we calculate the ADR of $x_2$ in (8) by  
\begin{eqnarray}
\begin{array}{l}\label{eqR2}
\mathbb{R}^{(2)}=\mathbb{R}_\textrm{BPSK}(\gamma_3), 
\end{array}
\end{eqnarray}
where $\gamma_3=\beta^2/\sigma_N^2$ is the SNR in the detection of $x_2$.

At last, the total ADR of the cocktail BPSK will be obtained by adding up the ADRs of $x_1$ and $x_2$, i.e., 
\begin{eqnarray}
\begin{array}{l}\label{eqR2}
\mathbb{R}=\mathbb{R}^{(1)}+ \mathbb{R}^{(2)}.
\end{array}
\end{eqnarray}

Different from the optimal detection in conventional modulations, the cocktail BPSK utilizes higher symbol energy
\begin{eqnarray}
\begin{array}{l}\label{eqR2}
E_1=(1/2)(\alpha+\beta)^2+(1/2)(\alpha-\beta)^2=\alpha^2+\beta^2=E_{in},
\end{array}
\end{eqnarray}
to detect $x_1$ of two possible amplitudes while utilizes lower symbol energy $E_2 =\beta^2$ to detect $x_2$.  Hence, the total symbols' energy utilized in the detection is $E_1+ E_2 =\alpha^2+2\beta^2$, which is larger than the input energy $E_{in}$ in (6).

To illustrate the performance for the detection of $x_1$ and $x_2$, we plot the numerical results of $\mathbb{R}^{(1)}$ and $\mathbb{R}^{(2)}$ versus linear SNR $E_{in}/\sigma_N^2$ in Fig.2(a), where both the ADRs improves with the increase in SNR.  However, the contributions from $\mathbb{R}^{(1)}$ and $\mathbb{R}^{(2)}$ to the ADR of cocktail BPSK can be adjusted by changing the ratio between $\alpha$ and $\beta$.  Comparing the cases of $\beta/\alpha = 0.3$ and $0.9$, we may find that $\mathbb{R}^{(1)}$ is not sensitive to this ratio in the low-SNR region.  As SNR increases, larger ratio of $\beta/\alpha$ will result in lower $\mathbb{R}^{(1)}$.  On the other hand, larger $\beta/\alpha$ always leads to higher $\mathbb{R}^{(2)}$, in both low- and high-SNR regions.  Thus, in the case of larger $\beta/\alpha$, the ADR of cocktail BPSK collects more contribution from $\mathbb{R}^{(2)}$ while suffers relatively small contribution from $\mathbb{R}^{(1)}$ in the low-SNR region.  It is interesting to note that the cocktail BPSK obtains higher gains in the cases of low SNRs and larger values of $\beta/\alpha$, which can be easily found by comparing with the channel capacity in the case of $\beta/\alpha = 0.9$ at low SNRs.
	
To express the curves in a traditional manner {\cite{Es2Eb}}, the linear SNR at the horizontal axis in Fig. 2(a) is replaced by the logarithmic ratio 
\begin{eqnarray} \label{EbEs}
\log_{10} \left(E_b/{\sigma_N^2}\right) = \log_{10} \left(\frac{E_{in}/{\sigma_N^2}}{\mathbb{R}}\right)
\end{eqnarray}
in decibels, where $E_b$ is the bit energy.  The results in a broader range of SNR are plotted in Fig.2(b), where the ADRs of cocktail BPSK are saturated at very high SNRs due to the limitation on the freedom degrees of modulated symbols.

\section{Comparison with the Channel Capacity}
In this section, the ADR of cocktail BPSK is compared with the channel capacity through theoretical analysis and numerical results. 

\subsection{Theoretical Analysis}
For a given input energy in (6), the ADR difference between cocktail BPSK and conventional BPSK is
\begin{eqnarray}
\begin{array}{l}\label{eqR2}
\begin{split}
\Delta\mathbb{R} &= \mathbb{R}-\mathbb{R}_{conv}\\& =\frac{1}{2}\mathbb{R}_\textrm{BPSK}(\gamma_1) + \frac{1}{2} \mathbb{R}_\textrm{BPSK}(\gamma_2) + \mathbb{R}_\textrm{BPSK}(\gamma_3) \\ & \qquad - \mathbb{R}_{conv}(E_{in}/ \sigma_N^2),
\end{split}
\end{array}
\end{eqnarray} 
where $\mathbb{R}_{conv}$ is the ADR of conventional BPSK, and the SNRs $\gamma_1=(\alpha+\beta)^2/ \sigma_N^2$, $\gamma_2=(\alpha-\beta)^2/ \sigma_N^2$, $\gamma_3=\beta^2/ \sigma_N^2$.

To sort out the approximation at zero SNR, we use \eqref{rateBPSK} in the appendix to replace $\mathbb{R}_{conv}$ and obtain 
\begin{eqnarray}\label{eqdC}
\Delta\mathbb{R} \approx \mathbb{R} - C = ({\log_2}e)\frac{\beta^2}{\sigma_N^2}>0
\end{eqnarray}
as $[E_{in}/\sigma_N^2]$ goes to zero, where $\Delta\mathbb{R}$ is the extra ADR with the cocktail BPSK beyond the channel capacity. 

%%%%%%%%%%
\begin{figure}[!t]
	\centering
	\subfigure[]{
		\includegraphics[width=0.5\textwidth]{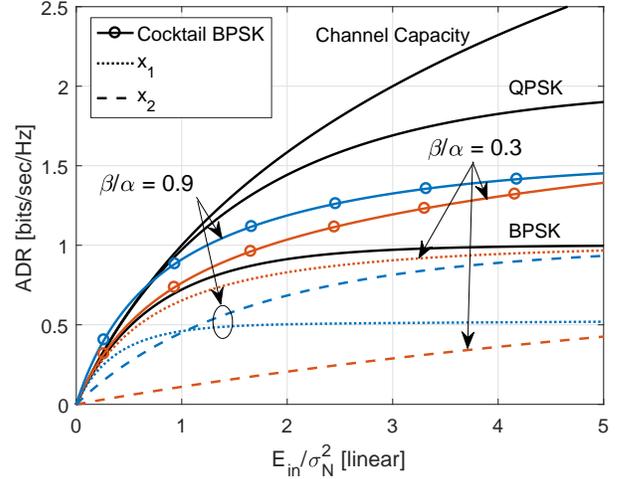}
		\label{fig2a}}
	\subfigure[]{
		\includegraphics[width=0.5\textwidth]{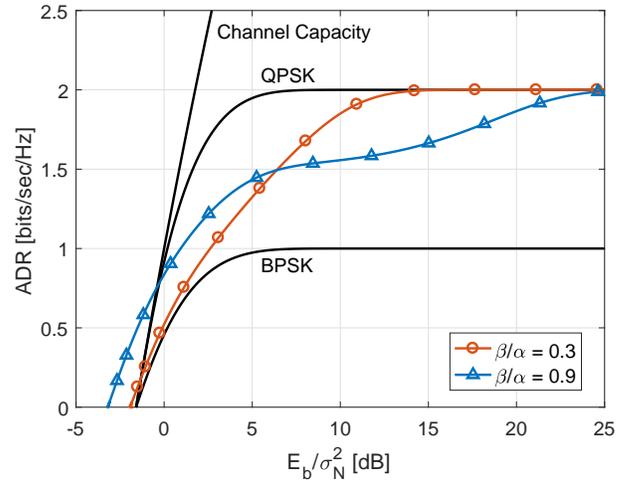}
		\label{fig2b}}
	\caption{ADR comparison between the cocktail BPSK and conventional BPSK over AWGN channels versus (a) linear $E_{in}/\sigma_N^2$ and (b) $E_b/\sigma_N^2$ in [dB].}
	\label{fig2}
\end{figure}

\begin{figure}[!t]
	\centering
	\subfigure[]{
		\includegraphics[width=0.45\textwidth]{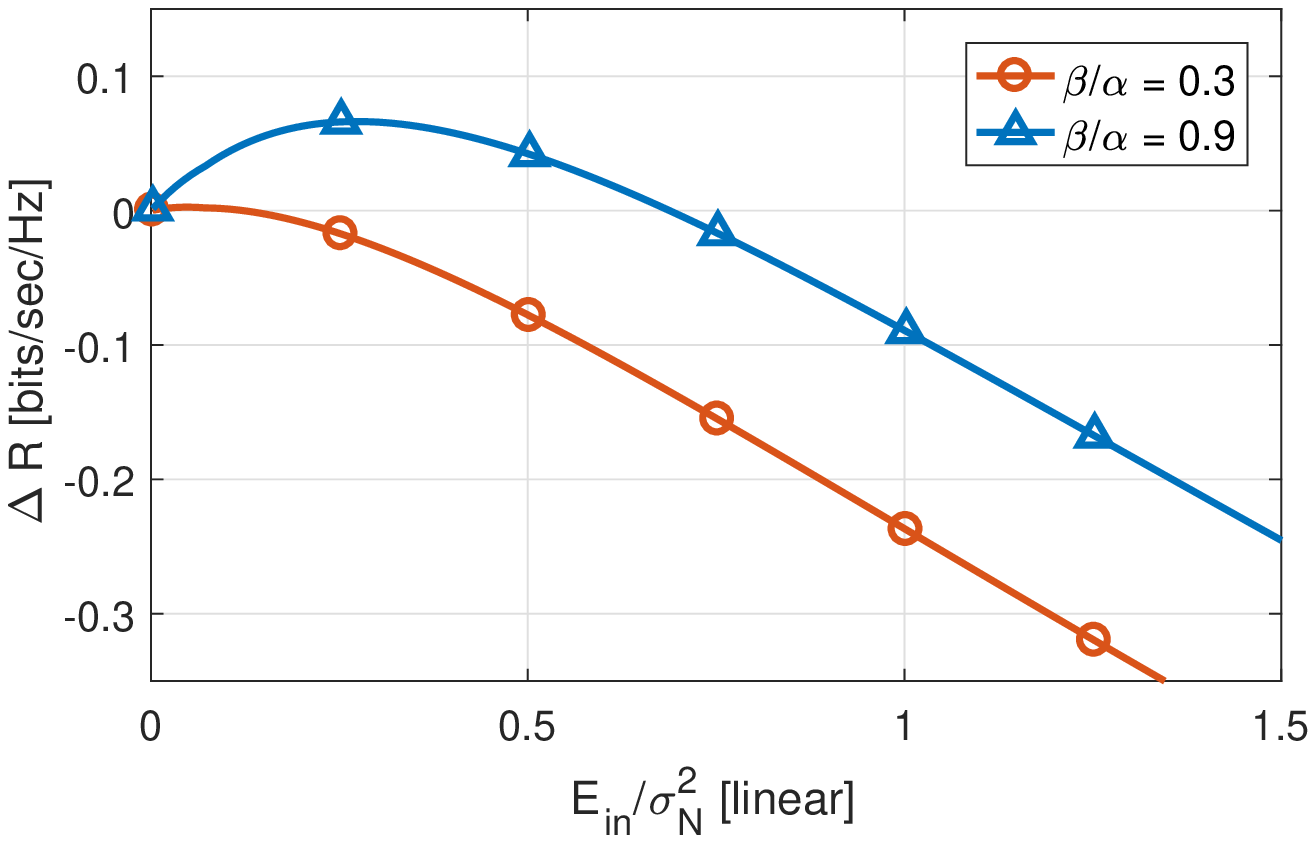}
		\label{fig3a}}
	\subfigure[]{
		\includegraphics[width=0.45\textwidth]{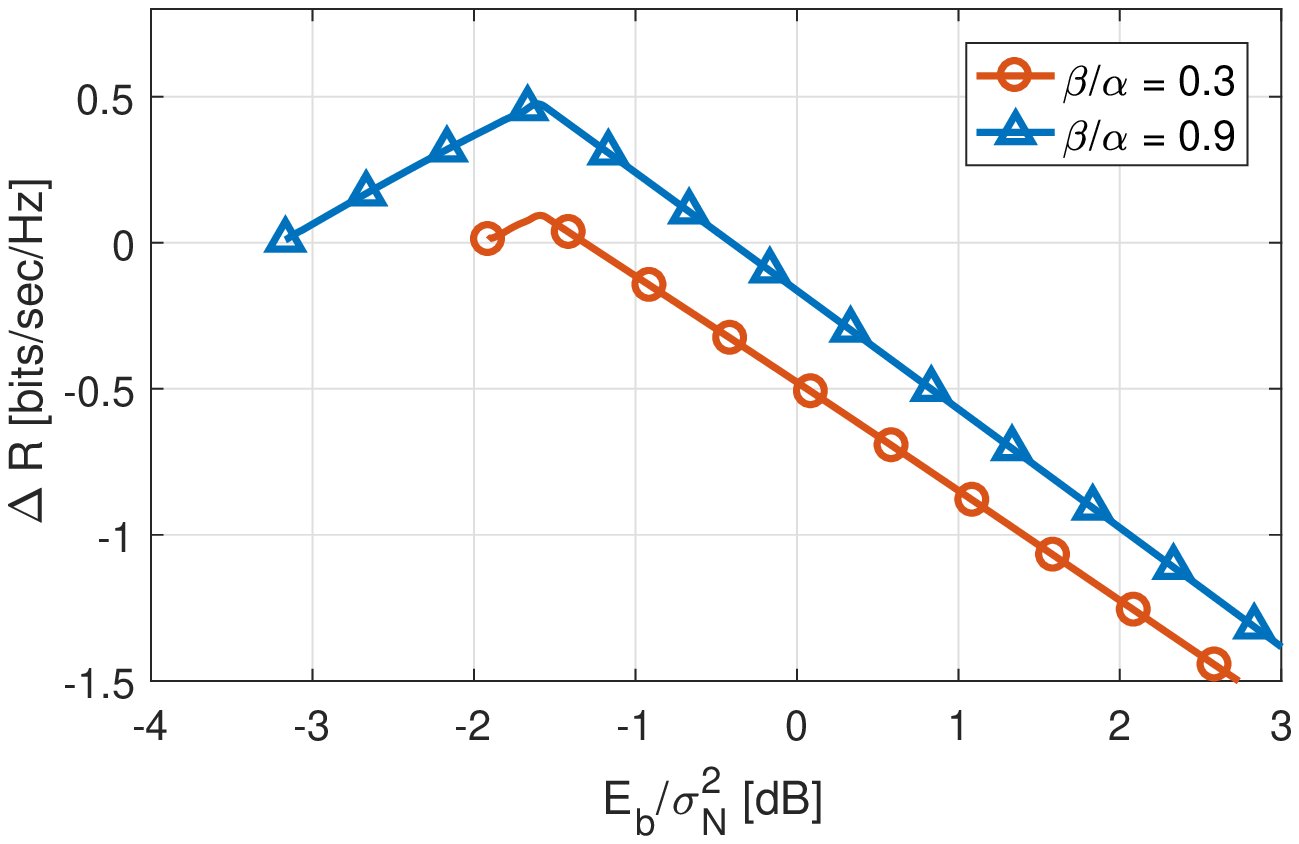}
		\label{fig3b}}
	\caption{ADR differences between the cocktail BPSK and the channel capacity versus (a) linear $E_{in}/\sigma_N^2$ and (b) $E_b/\sigma_N^2$ in [dB].}
	\label{fig3}
\end{figure}

\subsection{Numerical Results}

To provide a clear comparison between the ADR of the cocktail BPSK and the channel capacity, in Fig. 3(a) we plot the numerical results of $\Delta\mathbb{R} = \mathbb{R}-C$ in the low-SNR region, i.e., linear ratio $E_{in}/\sigma_N^2 \in [0, 1.5]$.  Moreover, the results versus logarithmic ratio of $E_b/{\sigma_N^2}$ are shown in Fig. 3(b).  In both figures, we can find the significant ADR gains over the channel capacity at low SNRs.  In addition, the larger $\beta/\alpha$ results in higher ADR gain with the cocktail BPSK, which agrees with the theoretical prediction in (15).   

However, the ADR gain of the cocktail BPSK gets smaller and even becomes a negative value as SNR increases.  The reason behind is that $\mathbb{R}^{(1)}$ withdraws its contribution and saturated first at high SNRs, as mentioned in Section II-B.  

%%%%%%%%%%%%%%%%%%%%   

\section{Conclusion}
In this paper, cocktail BPSK was proposed by layering two independent BPSK symbols in a parallel transmission.  In contrast to conventional signal processing, we translated the interference between the two BPSK symbols into the problem of one BPSK symbol having two equiprobable amplitudes.  Based on the concept of mutual information, the sum ADR of the two BPSK streams was found to achieve at higher values than the channel capacity at low SNRs, which was substantiated by both theoretical derivations and numerical results.

\appendix

To begin with, we calculate the ADR of conventional BPSK and express it as a function of SNR.

For a given amplitude $A$ of BPSK, its ADR is calculated using 
\begin{equation}\label{rBPSK1}
\begin{split}
& \mathbb{R}_\textrm{BPSK}(A,{\sigma_N^2}) = {\rm{H}}(Y) - {\rm{H}}(N) \\
& = \mathcal{E} \{- \log_2 p(y)\} - {\log _2}(\sqrt {2\pi e{\sigma_N^2}}) \\
& = - \int_{-\infty }^{ + \infty } {p(y){{\log }_2}p(y){\rm{d}}y} - {\log _2}(\sqrt {2\pi e{\sigma_N^2}} ),
\end{split}
\end{equation}
where $\mathcal{E} \{\cdot \}$ is the expectation operator, and ${\rm{H}}(Y) = $ is the entropy of the received BPSK signal with the probability density function given by
\begin{equation}
p(y) = \frac{1}{2}\frac{1}{\sqrt{2\pi \sigma_N^2}}\left(e^{-\frac{(y-A)^2}{2 \sigma_N^2}} + e^{-\frac{(y + A)^2}{2 \sigma_N^2}}\right).
\end{equation}
and ${\rm{H}}(N) = {\log _2} (\sqrt{2 \pi e \sigma_N^2})$ is the entropy of the AWGN.

To define $\gamma = A^2/\sigma_N^2$, \eqref{rBPSK1} is re-written as 
\begin{equation}
\begin{split}
\mathbb{R}_\textrm{BPSK}(A,\sigma_N^2) = 1 - \int_{-\infty}^{\infty} \frac{1}{\sqrt{\frac{16\pi A^2}{\sigma _N^2}}} & \exp \left(- \frac{\left(x - \frac{4A^2}{\sigma _N^2} \right)^2}{\frac{A^2}{2\sigma _N^2}} \right)\\
& \times \log_2 \left( 1 + e^{-x} \right) dx
\end{split}
\end{equation}

Subsequently, the ADR of conventional BPSK can be expressed by a function of the SNR $\gamma$ as  
\begin{eqnarray}
\begin{split}
\mathbb{R}_\textrm{BPSK}(\gamma)= 1 &- \frac{1}{{4\sqrt \pi  }}{\gamma ^{ - \frac{1}{2}}}{e^{ - \gamma }} \\
& \times \int_{-\infty}^{\infty} {{{\log }_2}\left( {1 + {e^{ - x}}} \right){e^{\frac{x}{2} - \frac{{{x^2}}}{{16\gamma }}}}} dx.
\end{split}
\end{eqnarray} 

Next, we will illustrate that the derivative of the ADR of the conventional BPSK is approximate to that of the channel capacity at low SNRs through its approximation at $\gamma=0$.  

The first-order derivative of $\mathbb{R}_\textrm{BPSK}(\gamma)$ is   
\begin{eqnarray}
\begin{aligned}
\frac{{d{\mathbb{R}_\textrm{BPSK}}(\gamma )}}{{d\gamma }} &= \frac{1}{{4\sqrt \pi  }}{\gamma ^{ - \frac{1}{2}}}{e^{ - \gamma }}\int_{-\infty }^{\infty } {{{\log }_2}\left( {1 + {e^{ - x}}} \right){e^{\frac{x}{2} - \frac{{{x^2}}}{{16\gamma }}}}} dx \\
& - \frac{1}{{4\sqrt \pi  }}{\gamma ^{ - \frac{1}{2}}}{e^{ - \gamma }}\int_{-\infty }^{\infty } {\frac{{{x^2}}}{{16{\gamma ^2}}}{{\log }_2}\left( {1 + {e^{ - x}}} \right){e^{\frac{x}{2} - \frac{{{x^2}}}{{16\gamma }}}}} dx \\
& + \frac{1}{{8\sqrt \pi  }}{\gamma ^{ - \frac{3}{2}}}{e^{ - \gamma }}\int_{-\infty }^{\infty } {{{\log }_2}\left( {1 + {e^{ - x}}} \right){e^{\frac{x}{2} - \frac{{{x^2}}}{{16\gamma }}}}} dx.
\end{aligned}
\end{eqnarray}

By drawing (20) in Fig. 4, we find that this derivative is equal to $\log_2 e$ at $\gamma = 0$.

Using Taylor expansion
\begin{equation}
\begin{array}{l}
f(x) \approx f(0) + (df/dx)x,  
\end{array}
\end{equation} 
for the positive value $x<<1$, to the channel capacity $C(\gamma)=\log_2(1+\gamma)$ and the ADR of conventional BPSK yields the approximation 
\begin{equation}\label{rateBPSK}
\mathbb{R}_\textrm{BPSK}(\gamma) \approx \mathbb{R}_\textrm{BPSK}^{'}(0)\gamma = C^{'}(0)\gamma = (\log_2 e)\gamma \approx C(\gamma),
\end{equation} 
where ${dC(\gamma )}/{d\gamma } = \log_2e$ at $\gamma=0$ is applied.
\begin{figure}[htb] 
	\centering
	\includegraphics[width=0.48\textwidth]{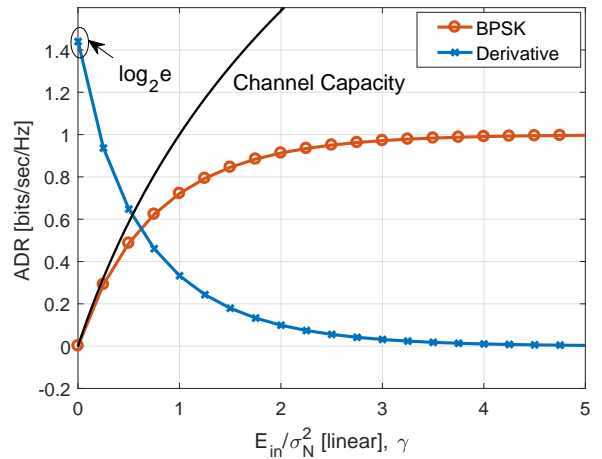}
	\caption{Numerical integration for the ADR of BPSK and its derivative.}
	\label{fig4}
\end{figure}

\end{document}